\documentclass[epj-spec]{svjour}
\usepackage{graphics}
\usepackage{xspace}
\usepackage{graphicx}
\newcommand{\pt}{\ensuremath{p_{T}}\xspace}

\newcommand{\pp}{\ensuremath{p+p}\xspace}

\newcommand{\ppbar}{\ensuremath{p+\bar{p}}\xspace}
\newcommand{\ee}{\ensuremath{e^{+}+e^{-}}\xspace}

\newcommand{\sqsRhic}{\ensuremath{\sqrt{s}=200}\xspace}

\newcommand{\cdf}{\ensuremath{\sqrt{s}=1.8}\xspace}

\newcommand{\dndxi}{\ensuremath{dN/d\xi}\xspace}
\newcommand{\xid}{\ensuremath{\xi}\xspace}
\newcommand{\xiz}{\ensuremath{\xi_{0}}\xspace}

\newcommand{\lam}{\ensuremath{\Lambda}\xspace}

\newcommand{\Ks}{\ensuremath{\mathrm{K^{0}_{S}}}\xspace}

\begin{document}
\title{Systematic study of the jet fragmentation function for inclusive
jet-production in $p+p$ collisions at $\sqrt{s}=200$ GeV in STAR}

\author{Mark Heinz\inst{1}\fnmsep\thanks{\email{mark.heinz@yale.edu}} for the STAR Collaboration }
\institute{Yale University, Physics Department, WNSL, 272 Whitney
Ave, New Haven, CT 06520, USA}
\abstract{ Jet fragmentation functions measured in $e^+e^-$ and
$p+\bar{p}$ experiments are well-described on an inclusive hadron
level by QCD-based calculations. Fragmentation is expected to be
modified by the presence of a strongly interacting medium, but full
theoretical description of this modification must still be
developed. It has recently been suggested that particle-identified
fragmentation functions may provide additional insight into the
processes underlying jet quenching. To assess the applicability of
QCD-based fragmentation calculations to RHIC data, and to provide a
baseline with which to compare fragmentation function measurements
in heavy ion collisions, we present the first measurements of
charged hadron and particle-identified fragmentation functions of
jets reconstructed via a midpoint-cone algorithm from p+p collisions
at 200 GeV in STAR. We study the dependence on jet cone-size and
jet-energy, and compare the results to PYTHIA simulations based on
the Modified Leading Log Approximation (MLLA).
} %end of abstract
\maketitle
\section{Introduction}\label{intro}

The measurement of jets in \pp and \ee collisions provides a
stringent test of perturbative QCD and factorization. In particular
there have been recent efforts to theoretically describe the hadron
multiplicity and its momentum distribution in jets in terms of
fragmentation functions \dndxi, where
$\xi=ln(1/x)=ln(E_{jet}/p_{h})$. The modified leading logarithmic
approximation (MLLA) describes the parton shower and has been
successful at analytically computing the momentum and angular
distributions of partons in jets \cite{MLLA}. Assuming local parton
hadron duality (LPDH) these distributions are converted to hadron
fragmentation functions \dndxi using a proportionality factor of
order 1. Monte Carlo parton shower, such as implemented in PYTHIA,
are also based on MLLA. Recent comparisons of the experimentally
measured \dndxi distribution of di-jets in \cdf TeV \ppbar collision
at CDF to MLLA exhibit very good agreement \cite{CDF:2003}.

The measurement at RHIC of the high-pt particle suppression in
inclusive hadrons ($R_{AA}$) and the disappearance of back-to-back
correlations in Au-Au collisions is commonly interpreted as medium
induced jet-quenching
\cite{STAR-HighPt-Raa:2003,STAR-Dihadron:2003}. However both of
these leading particle observables are from a biased sample of jets.
Therefore in order to access jets in an unbiased way and to probe
the underlying parton kinematics we need to perform full
jet-reconstruction. A wide variety of jet-finding algorithms are now
available and tested \cite{Fastjet}. The jet fragmentation function,
expressed in terms of \xid, also referred to as ``hump-back
plateau", is a sensitive observable for the medium-modification of
jets \cite{Borghini-Wiedemann:2005}. The \dndxi distribution has the
advantage of highlighting the parton energy redistribution within
jets from harder to softer partons believed to be the predominant
effect of gluon radiation. Some numerical results for \dndxi for
different jet-energies and jet-cone opening angles can be found here
\cite{Sapeta-Wiedemann:2007}. In that paper the authors also predict
particle species dependent fragmentation functions and the change of
particle ratios in jets as a function of jet-energy and \pt.

Thus the goal of the present analysis of STAR \pp data at 200 GeV is
threefold. First, we want to confirm that the MLLA framework also
works at much lower jet-energies of the order of 15-50 GeV such as
produced at RHIC. For this we compare \dndxi distributions for
charged hadrons in jets of various cone-sizes to predictions by MLLA
as implemented in the PYTHIA Monte Carlo program (version 6.4).
Second, we will present the first measurement of particle-identified
\dndxi distributions for \lam and \Ks particles in \pp collisions.
These will be usefull to test and constrain the parameters in the
Sapeta-Wiedemann model. And third, the \dndxi distributions will be
used as a baseline for future comparisons to medium modified \dndxi
from jets in heavy ion collisions at \sqsRhic GeV. First progress
towards this goal has been made recently \cite{STAR-Joern-HP08}.

\section{Analysis Technique}\label{analysis}

\subsection{Event Selection and Online triggers}\label{triggers}
The data-sample presented here was taken during the 2005/2006
polarized pp running at RHIC. The total sampled luminosity was 6.2
$pb^{-1}$, which resulted in 8.3 million usable events. The two main
online triggers used for the jet-analysis were the High-Tower (HT)
and the Jet-Patch (JP) trigger. The tower granularity of the STAR
barrel electromagnetic calorimeter (BEMC) is $\Delta\eta \times
\Delta\phi = 0.05 \times 0.05$ and it has a coverage of $-1< \eta
<1$ and $\Delta\phi=2\pi$. The analysis presented here used the JP
trigger, requiring the energy of an area of $\Delta\eta \times
\Delta\phi = 1 \times 1$ to be above a threshold of 8 GeV. The
reason for preferably using a JP trigger (instead of a HT) is that
the trigger bias towards leading neutral particles ($\pi^0$) is less
pronounced when integrating energy over a larger BEMC area, instead
of a single tower. More details about the various BEMC triggers and
their biases can be found in \cite{STAR-Spin:2008}.
\begin{figure}[b]
\includegraphics[width=0.49 \textwidth]{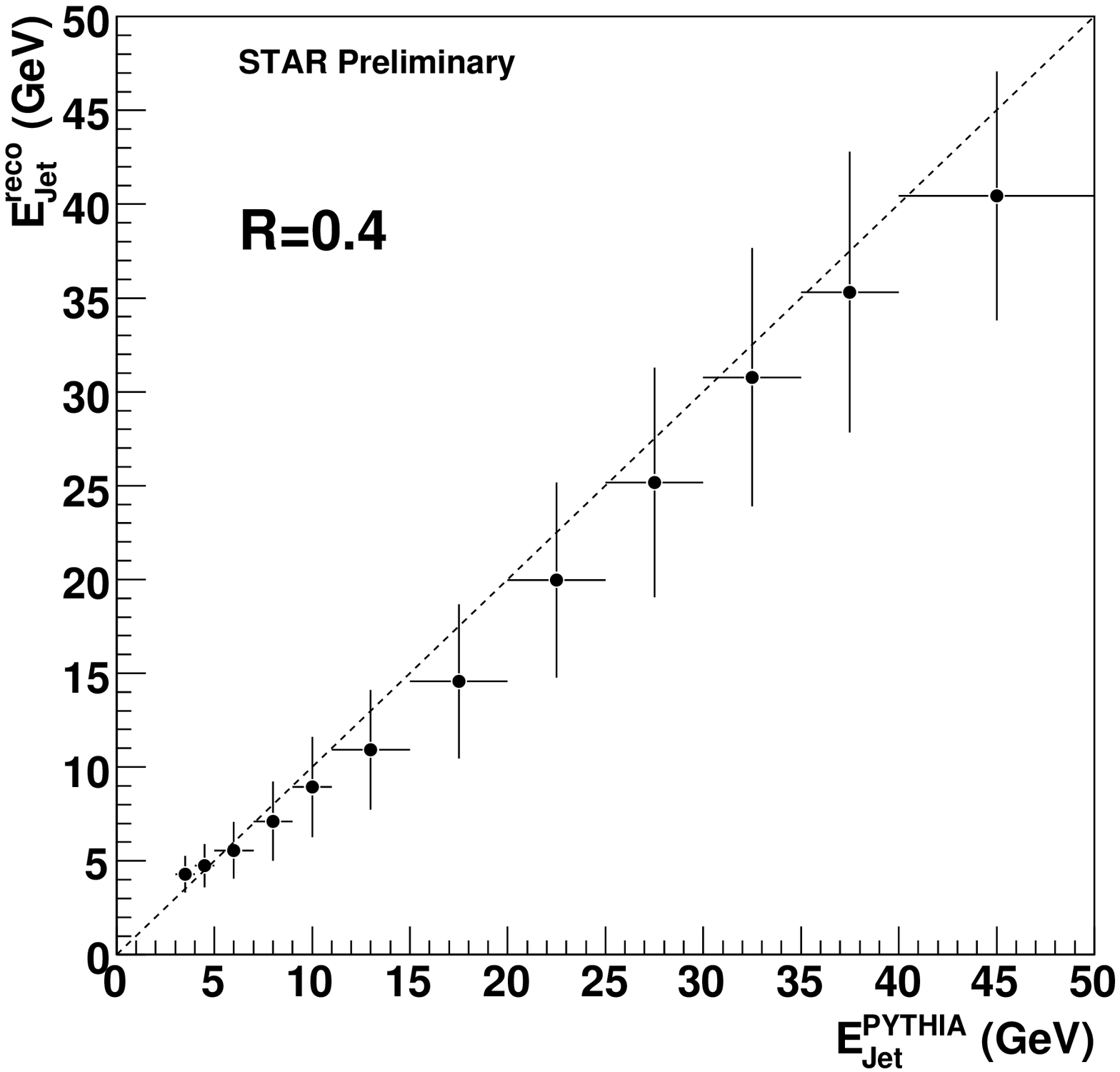}
\includegraphics[width=0.49 \textwidth]{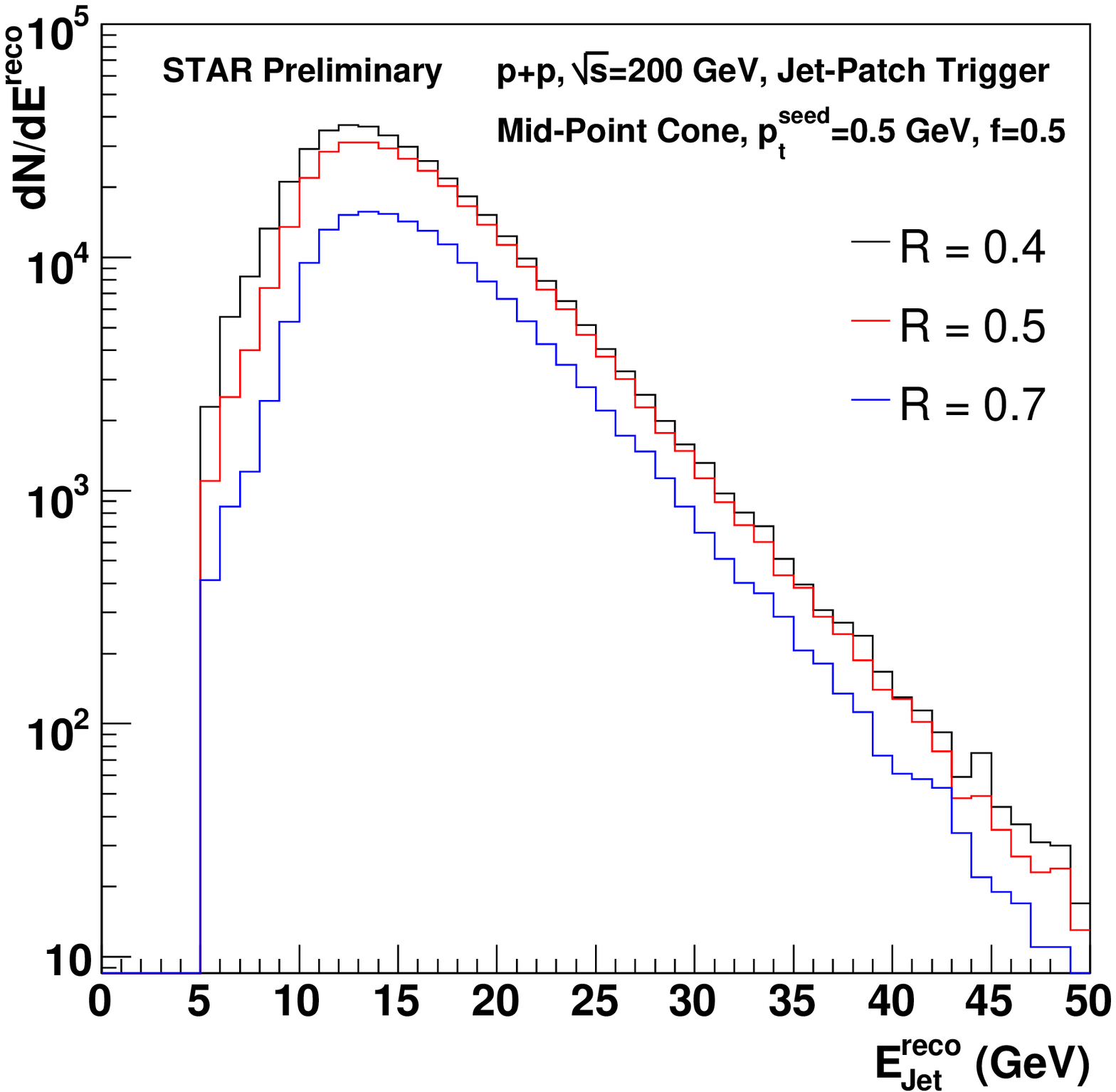}
\caption{(Left) Energy resolution as a function of PYTHIA jet-energy
for the mid-point cone algorithm with radius R=0.4. The y-axis is
the reconstructed jet-energy after GEANT and track reconstruction.
(Right) Reconstructed leading jet-energy spectrum for the highest
jet in the event. Differences in magnitude are due to the varying
acceptance as a function of cone-radius} \label{fig:EnerReso}
\end{figure}

\subsection{Jet-Finding algorithm}\label{jetfinding}
For this study we used the midpoint-cone jet-finding algorithm
commonly used in \ppbar collisions at CDF. This has also been used
in previous STAR publications and the measured jet cross-section
agrees well with NLO pQCD \cite{STAR-Spin:2006}. The main
jet-reconstruction parameters were set to be $R=0.4-0.7,
p_{T}^{seed}=0.5$GeV/c and $f_{split}^{merge}=0.5$. In addition we
only accepted jets whose Neutral Energy Fraction (NEF) was within
$0.05< NEF < 0.85$ to avoid a large trigger bias and to remove
backgrounds with large NEF. Also the jet-axis was required to be
within a given $\eta$ range (dependent on R) to ensure full
acceptance for the cone. For this analysis only the leading jet in
each event was considered.

In order to compare our fragmentation function results to theory we
used the PYTHIA Monte Carlo simulation (version 6.4)
\cite{Pythia64}. The PYTHIA output was further processed through our
GEANT detector response simulator, and then analyzed in exactly the
same way as the data. The PYTHIA sample was composed of several
pt-hard bins which were weighted according to their cross-section.
We applied the same jet-finding algorithm, with the same parameters
to find jets and then extract \dndxi distributions for charged
hadrons.

In figure \ref{fig:EnerReso} (left) we studied the energy resolution
by comparing the reconstructed jet-energy, after GEANT, tracking and
jet-finding, with the energy obtained from applying the algorithm to
the PYTHIA input. An approximately constant energy resolution of
$\sim25\%$ was obtained for all measurable jet-energies. On the
right panel of figure \ref{fig:EnerReso} we show the measured
jet-energy spectrum for the leading jets in each event. In this
study the reconstructed energy was not corrected for the shift with
respect to the input energies. The statistical reach allows to
reconstruct about 500 jets in the 40-50 GeV energy range.
\begin{figure}[b]
\includegraphics[width=\textwidth]{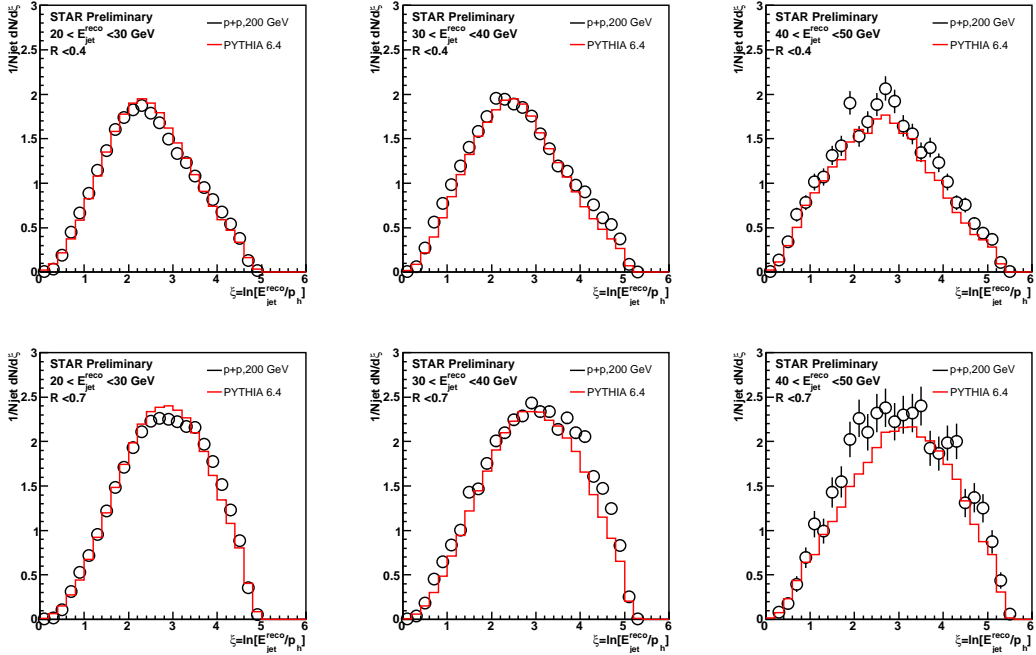}
\caption{Charge particle \dndxi distribution for reconstructed
jet-energies 20-50 GeV (leading jets only) and cone-radii
R=0.4(upper) and R=0.7(lower). Only statistical errors shown.}
\label{fig:ChrgXi}
\end{figure}
In addition jet-energy was corrected for two effects. In jets with
identified electrons, whose energy would be double-counted due to
them being measured by both the Time Projection Chamber (TPC) and
BEMC, only the track momentum was used. Second, for charged hadrons
in jets which projected to an active BEMC tower (deposited energy
above pedestal) the expected energy deposit of a minimum ionizing
particle (MIP) was subtracted from the tower energy.

\subsection{Particle Identification}\label{PID}
We used identified V0 particles, \Ks and \lam, via their dominant
hadronic decay to charged particles. These were chosen because STAR
has excellent efficiency and particle identification over a large
kinematical range for reconstructing these decays ($1 < \pt$(GeV/c)$
< 8(\lam),10(\Ks)$). Using the methods described in
\cite{STAR-Strange-PPpaper} we reconstructed the invariant mass
distributions of V0's and identify \Ks and \lam particles. These
were then added to the particle pool and their charged daughters
were removed before the jet-finding step. This resulted in a total
of $\sim$ 100k jets with \Ks and $\sim$ 45k jets with \lam. The
$\xi$ distributions for \lam and \Ks were corrected for
reconstruction efficiency using a sample of simulated jets. For \pt
greater than 2.5 GeV/c the reconstruction efficiencies were
independent of \pt and of the order of 30\% and 15\% for \Ks and
\lam respectively.
\begin{figure}[t]
\includegraphics[width=\textwidth]{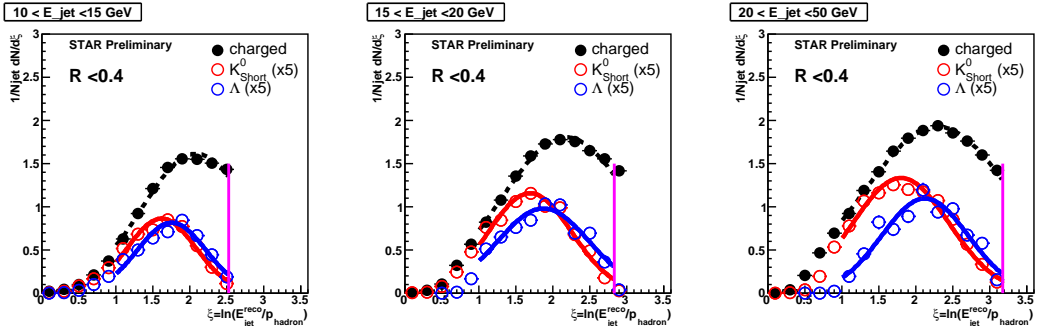}
\caption{\dndxi for identified strange particles in jets for three
different jet-energies. Fits are Gaussian functions and are only
used to obtain the peak value $\xi_{0}$. The vertical (purple) line
indicates the \pt-cutoff of 1 GeV/c for good identification of
strange particles. Charged hadrons are shown for comparison. Only
statistical errors shown.} \label{fig:V0Xi}
\end{figure}
\begin{figure}[b]
\centering
\includegraphics[width=0.5 \textwidth]{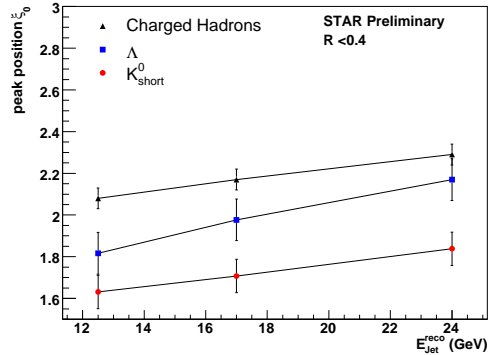}
\caption{Peak value $\xi_{0}$ of \dndxi distribution for charged
hadrons, \Ks and \lam for R=0.4 and different jet-energy selections.
The peak values were estimated from Gaussian fits within the range
of $\xi>1$ and $\pt>1$ GeV/c as shown on figure \ref{fig:V0Xi}.}
\label{fig:meanXi}
\end{figure}
\section{Results and Discussion}\label{results}

In figure \ref{fig:ChrgXi} we compare the results of charged hadron
\dndxi distribution to results from PYTHIA, for two different
cone-sizes and three jet-energy bins. The overall agreement between
data and model is rather good although small deviations can be
found, in particular for larger cone-sizes. The fact that the
differences between experiment and model are small for R=0.4 and
more significant for R=0.7, could be evidence of the importance of
next-to-leading order corrections to PYTHIA since it is expected
they are more important at larger angles with respect to the
jet-axis.

In figure \ref{fig:V0Xi} we show the results for identified \Ks and
\lam particles for different jet-energies, and as a reference the
charged \dndxi distribution. The jet-energy bins were chosen to
optimize the available statistics of \Ks and \lam particles. \dndxi
distributions are expected be Gaussian around their peak and
therefore we used a Gaussian fit in order to extract the peak value
\xiz. According to perturbative QCD \xiz is expected to be ordered
with particle mass \cite{Dokshitzer:1991}. The result of extracting
\xiz for different particle species and jet-energies is shown in
figure \ref{fig:meanXi}. The data seem to indicate that the QCD
predicted mass ordering is not obeyed since the peak value \xiz for
\lam is equal or higher than \xiz for \Ks for all three energies.
However the significance of this result needs to be confirmed once
the systematic errors have been evaluated. A similar observation of
violated mass ordering between pions, protons and kaons has been
seen by a range of \ee experiments at $\sqrt{s}$=10-100 GeV
\cite{Babar-Anulli:2004}. For comparison to this result,
measurements in \ee collisions at $\sqrt{s}$=10.54 GeV from BABAR
show the peak of the kaon and proton distribution to be equal at
$\xiz \sim 1.6$.

\begin{figure}[t]
\includegraphics[width=0.49 \textwidth]{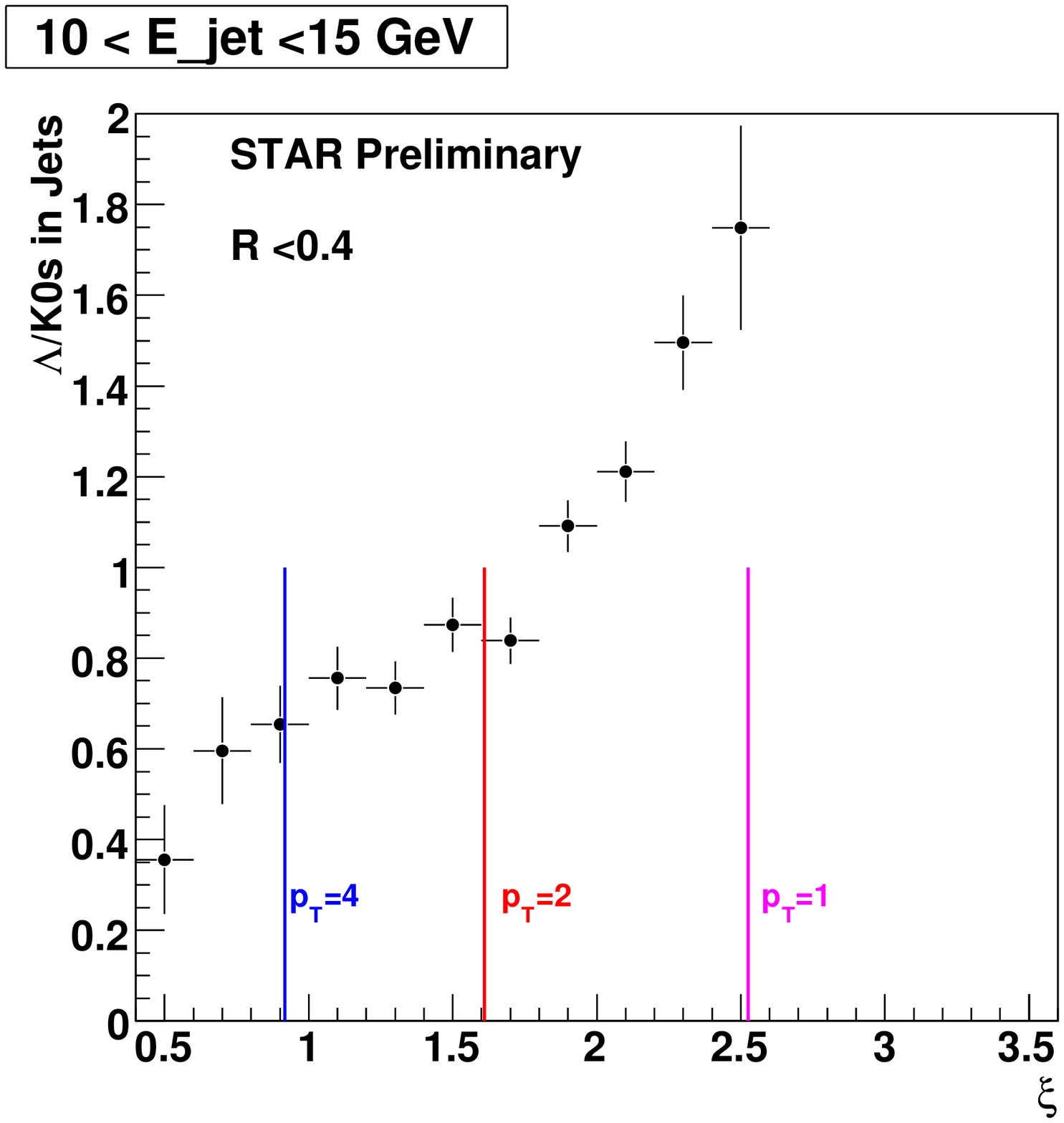}
\includegraphics[width=0.49 \textwidth]{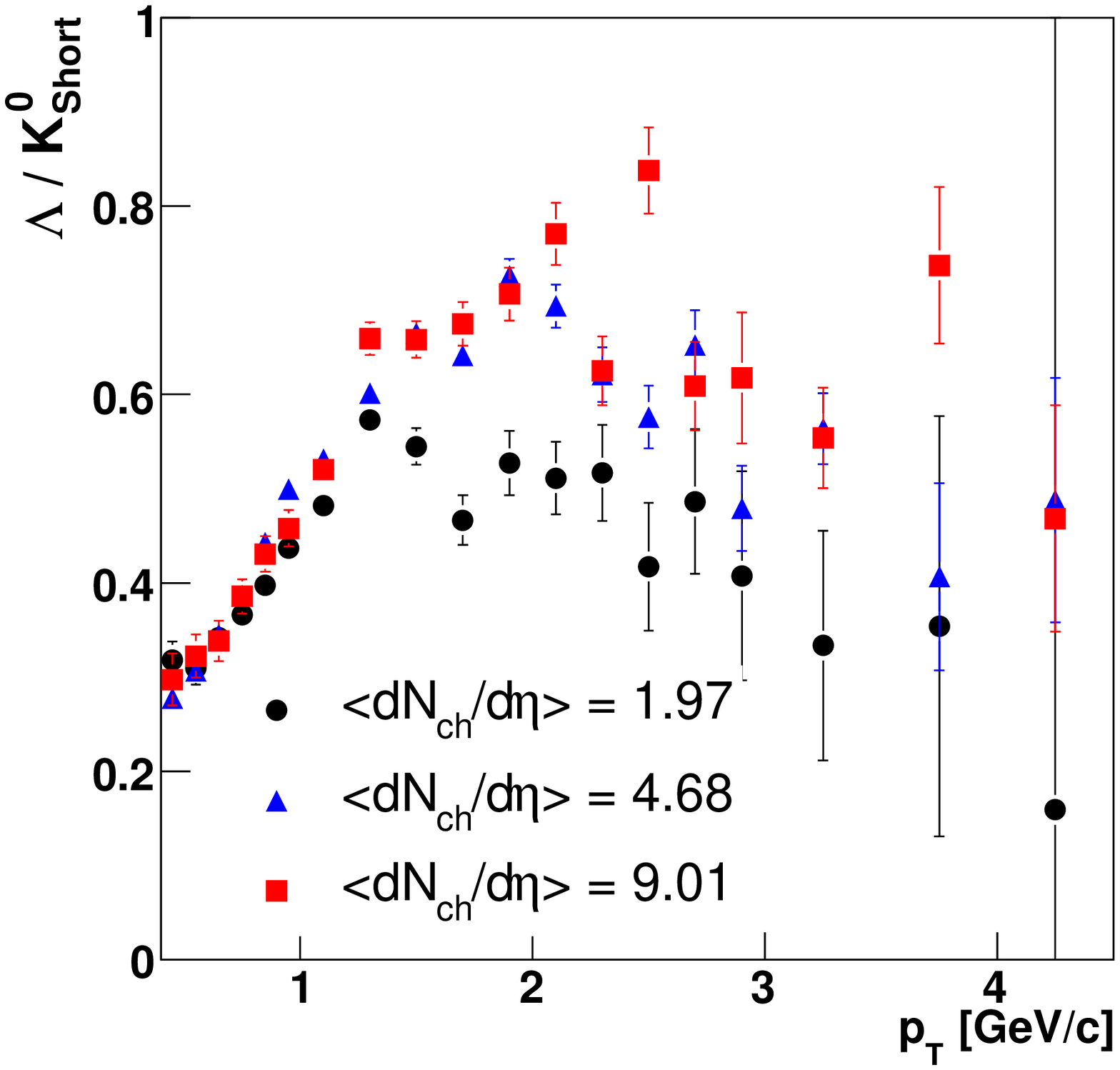}
\caption{(Left) \lam/\Ks ratio measured in jets as a function of
\xid. Vertical lines allow easier reference to \pt=1,2,4 $GeV/c$
labeled in the figure. Statistical errors only.(Right) The \lam/\Ks
ratio measured in inclusive p+p collisions as a function of event
multiplicity. From \cite{STAR-Strange-PPpaper}. } \label{fig:LamK0}
\end{figure}

In figure \ref{fig:LamK0} we study the relative production of \lam
vs \Ks in jets as a function of \xid. We can compare these results
to previous measurements of inclusive strange particle production in
p+p collisions at \sqsRhic (right panel)
\cite{STAR-Strange-PPpaper}. The values of the \lam/\Ks ratio at low
\xid (left panel) are consistent with the measurements at high-\pt
in the right panel. However the value of the ratio at high-\xid is
much larger than the value at a comparable \pt in inclusive
production. It grows significantly larger than 1 for $\xid >2$ and
does not exhibit the typical maximum as seen in the baryon-to-meson
ratio in inclusive production. Such a large baryon-to-meson ratio in
jets could be indicative of a different hadronization mechanism for
low momentum baryons.

In summary, we have shown the first measurement of charged particle
\dndxi distributions in inclusive jets in \pp collisions at \sqsRhic
at RHIC and concluded that they agree well with PYTHIA for small
cone-radii. Furthermore we have measured identified particle \dndxi
and observe that the peak values do not follow a particle mass
ordering scheme. In addition we observe a large strange
baryon-to-meson ratio ($>1$) at high \xid in jets. This apparent
disconnect between theory and experiment points to a more
fundamental lack of understanding of how baryons are produced. With
the start of the LHC (Large Hadron Collider) further measurements of
these fragmentation functions in vacuum and the medium should become
available and may help resolve these issues.

\end{document}